\documentclass[12pt]{article}
\usepackage[letterpaper]{geometry}
\usepackage[parfill]{parskip}
\usepackage{amsmath,amsthm,amssymb,bbm,amsfonts,dsfont}
\usepackage{mathtools}
\usepackage{cases}
\usepackage{graphicx}
\usepackage{tabularx}
\usepackage{microtype}
\usepackage{enumitem}
\usepackage{subfigure}			
\usepackage{xcolor}

\usepackage{authblk}

\usepackage{url}
\usepackage[colorlinks,citecolor=blue,urlcolor=blue,linkcolor=blue,linktocpage=true]{hyperref}
\usepackage{cleveref}
\crefformat{equation}{(#2#1#3)}
\crefrangeformat{equation}{(#3#1#4) to~(#5#2#6)}
\crefname{equation}{}{}
\Crefname{equation}{}{}

\usepackage[authoryear]{natbib}
\usepackage{multirow}

\newtheoremstyle{mythmstyle}
  {8 pt} 
  {3 pt} 
  {} 
  {} 
  {\bfseries} 
  {.} 
  {.5em} 
  {} 

\theoremstyle{mythmstyle}

\newtheorem{theorem}{Theorem}[section]

\newtheorem{proposition}[theorem]{Proposition}

\newtheorem*{example*}{Example}

\newtheorem*{definition*}{Definition}
\newtheorem{assumption}{Assumption}
\newtheorem*{remark*}{Remark}
\crefname{definition}{\textbf{definition}}{definitions}
\Crefname{definition}{Definition}{Definitions}
\crefname{assumption}{\textbf{assumption}}{assumptions}
\Crefname{assumption}{Assumption}{Assumptions}

\newcommand{\rss}{\vspace{-.4cm}}

\def\var{{\rm var}}

\newcommand{\bc}{\begin{center}}
\newcommand{\ec}{\end{center}}

\title{Least Squares Inference for Data with Network Dependency}
\author{Jing Lei, Kehui Chen, and Haeun Moon}
\begin{document}

\pagenumbering{Alph}
\begin{titlepage}
\thispagestyle{empty}
\maketitle

\begin{center}
  \textbf{Abstract}
\end{center}
We address the inference problem concerning regression coefficients in a classical linear regression model using least squares estimates. The analysis is conducted under circumstances where network dependency exists across units in the sample. Neglecting the dependency among observations may lead to biased estimation of the asymptotic variance and often inflates the Type I error in coefficient inference. In this paper, we first establish a central limit theorem for the ordinary least squares estimate, with a verifiable dependence condition alongside corresponding neighborhood growth conditions. Subsequently, we propose a consistent estimator for the asymptotic variance of the estimated coefficients, which employs a data-driven method to balance the bias-variance trade-off. We find that the optimal tuning depends on the linear hypothesis under consideration and must be chosen adaptively. The presented theory and methods are illustrated and supported by numerical experiments and a data example. 

\vspace*{.1in}


\end{titlepage}

\pagenumbering{arabic} 
\section{Introduction}\label{sec:introduction}

The analysis of dependent data with specific structures such as longitudinal data, times series data and spatial-temporal data constitutes long-standing research areas with significant theoretical and practical implications. Recently, data with complex interpersonal dependency characterized by networks are becoming increasingly available. For example, in some survey studies of school students, in addition to traditional variables collected on each unit such as drug use, smoking and mental health status, a friendship network among students may also be collected. The response of interest such as drug use is likely to have  dependency across units through friendship networks. In animal studies, the animal samples in their habitats naturally interact with each other. In addition to traditional individual data collection, animal researchers are now able to record a social contacting network through camera tracking combined with computer software such as idTracker \citep{perez2014idtracker}. The analysis of this type of data calls for statistical inference tools and theories that take into account the network dependency. In this work, we focus on developing methods and theory for the inference of regression coefficients, when there is network dependency across units of sample.  

We consider a regression setting $Y = X\beta+\epsilon$, where $Y\in\mathbb R^{n}$, $X\in\mathbb R^{n\times p}$ and $\epsilon \in \mathbb R^{n}$. Each sample point $1\leq i\leq n$ also corresponds to a node in a network.  Assume that we observe a network adjacency matrix $A$, where $A_{ij} = 1$ means an edge exists between individual $i$ and $j$ and $A_{ij}=0$ otherwise. We allow the entries of the error vector $\epsilon$ to be correlated through the network structure $A$. More specifically, a general dependence structure to consider is that the dependence between $\epsilon_i$ and $\epsilon_j$ decays as their graph distance $d(i,j)$, such as the shortest path distance, increases. Network auto-regressive error models and network moving-average error models both fall into this framework, where in the latter case dependency only spans a neighborhood of finite radius measured in network distance. This $Y = X\beta+\epsilon$ setting is particularly suitable if the main interest is in the estimation and inference of $\beta$ instead of modeling social network effects. For example, researchers in mental health studies are often interested in identifying biomarkers. A linear regression model with least squares estimate is arguably the most widely used approach and many of the tools developed for model interpretation, model selection, and multiple comparison correction are based on a linear model or a generalized linear model. Meanwhile, people are aware that neglecting the dependency among observations may result in an incorrect estimation of standard errors, and often inflates the Type I error in the inference of $\beta$. This has been well observed in data with temporal/spatial dependency \citep[][Ch. 5.3]{gaetan2010spatial}, and has recently been shown in data with network dependency \citep{lee2021network}. Therefore, in this work we focus on developing valid inference of $\beta$ based on the ordinary least squares estimate $\hat\beta$, where the estimation of $\var(\hat\beta)$ takes into account the network dependency but does not impose strong or parametric assumptions on the dependence structure. 

While there exists a large body of literature on the analysis and characterization of the network/graph itself, there is only limited work that addresses the fundamental normal approximation for the sum of random variables observed with network/graph dependence. Under the finite range dependence structure, central limit theorems can be established using tools developed in \cite{baldi1989normal,chen2004normal}, and some recent work studied CLT for infinite range network dependency \citep{shashkin2010berry, ogburn2017causal,kojevnikov2021limit}. The challenges of dealing with network dependency are at least twofold. The first challenge is the infinite dimensionality, which can be understood through the notion of network neighborhood growth.  When the average number of neighbors grows polynomially with the neighborhood radius measured by a graph distance, the graph can be embedded in a Euclidean lattice and theories for Euclidean spaces can be borrowed. Time series data and spatial data can be viewed as data on Euclidean lattices, and are special cases of low-dimensional embeddable graphs. A general network may have a faster neighborhood growth beyond the Euclidean lattice case and corresponds to a much richer and more complex embedding space \citep{weber2020neighborhood}.  The second challenge is the node heterogeneity, which means each node may have its own neighborhood growth sequence. Such a lack of symmetry makes it hard to carry over the useful tools developed for Euclidean lattices such as stationarity and mixing.

In this paper we make three theoretical and methodological contributions to least squares regression inference with general network dependence. First, we establish a central limit theorem for the ordinary least squares estimate $\hat\beta$, with a verifiable dependence condition alongside corresponding neighborhood growth conditions. Second, we propose a consistent estimator for the asymptotic variance of $\hat\beta$. The key quantity to estimate is the matrix $X^T\Sigma X$, where $\Sigma$ denotes the unknown covariance matrix of the noise vector $\epsilon$.
Our estimator $\widehat{X^T\Sigma X}$ uses the pair-wise products of fitted residuals to approximate the entries of $\Sigma$ and truncates a pair-wise product to $0$ if the graph distance between the two nodes exceeds a threshold $m$. It can be viewed as an adaptation of the sandwich estimator \citep{huber1967under, white1980heteroskedasticity,liang1986longitudinal} in the network dependency setting. Third, we propose a data-driven approach to determine the truncation order $m$ in coefficient inferences. We demonstrate that the optimal truncation order $m$ depends on the linear hypothesis under consideration, and must be chosen adaptively. In Section 2, we present the main theoretical results. In Section 3, we describe the data-driven estimate for the variance components. Numerical experiments and data examples are presented in Section 4 and 5, followed by a short discussion in Section 6. Proofs are included in the Appendix.

\rss\section{Inference of regression coefficients under network dependency}\rss
Consider a linear regression model written in matrix form
$$
Y=X\beta+\epsilon\,,
$$
where $Y\in\mathbb R^{n} = [y_1, \dots y_n]^T$, $X\in\mathbb R^{n\times p}=[x_{1\cdot},...,x_{n\cdot}]^T=[x_{\cdot1},...,x_{\cdot p}]$, $\beta\in\mathbb R^p$, and $\epsilon\in\mathbb R^n$  satisfying $\mathbb E(\epsilon|X)=0$.

We are interested in the case where the entries of $\epsilon$ may be dependent, and the strength of pairwise dependence can be quantified by the pairwise distances. 
Our general theory and methods work for all distances satisfying triangle inequality. In this paper our examples will focus on the shortest path distance defined on a network. To facilitate discussion, we assume that each sample point $i$ is a node on a graph denoted by a symmetric binary adjacency matrix $A$. Let $d(\cdot,\cdot)$ denote the shortest path distance, or other distance under consideration, between a pair of nodes.


Our goal is to establish asymptotic normality of the least squares estimate of $\beta$ under suitable conditions of the dependence decay and the network neighborhood growth.  The parameter dimensionality $p$ is assumed fixed.  In order to present the asymptotic theory, we adopt a fixed-design setting and consider a sequence of data generating models $(X,\Sigma,A)$ indexed by the sample size $n$, where for each $n$
\begin{itemize}
	\item $X$ is an $n\times p$ non-random design matrix;
	\item $A$ is an $n\times n$ symmetric binary adjacency matrix;
	\item $\Sigma$ is an $n\times n$ covariance matrix such that $\Sigma=\mathbb E \epsilon\epsilon^T$.
\end{itemize}
For notational simplicity, we drop the dependence on $n$ in $(X,A,\Sigma)$ and other related objects such as $Y$ and $\epsilon$.

We group our technical assumptions into three parts.  The first part is some general boundedness and moment conditions.
\begin{assumption}\label{ass:bound}
	There exists a constant $c$ such that
	\begin{enumerate}
	 	\item [(a)] $|x_{ij}|\le c$ for all $i\ge 1$, $1\le j\le p$;
	 	\item [(b)] $\sup_i\|\epsilon_i\|_4\le c$;
	 	\item [(c)] $c^{-1}I \preceq X^T\Sigma X/n \preceq c I$.
	 \end{enumerate}
\end{assumption}
\Cref{ass:bound} part (a) and part (b) are standard boundedness and moment conditions for the design matrix and noise variable, respectively.  The constant $c$ in these two parts can be relaxed to grow slowly with $n$, as long as they are dominated by other vanishing terms in the error bound.  \Cref{ass:bound} part (c) is much less stringent than requiring $c_1 I \preceq \Sigma \preceq c_2 I$, because it only requires the eigenvalues of $\Sigma$ after projecting onto the column subspace of $X$ to be bounded from above and away from zero.

The second part of assumptions is about the dependence decay among coordinates of $\epsilon$ when the graph distance grows.
\begin{assumption}[Correlation decay]\label{ass:decay}
	The process $(\epsilon_i:i\ge 1)$ satisfies an exponential correlation decay with respect to a distance measure on the index set $d(\cdot,\cdot):\mathbb N^2\mapsto \mathbb R^+\cup \{0\}$, such that the following hold for some constants $\rho\in(0,1)$ and $c>0$
	\begin{enumerate}
		\item [(a)]
		 $|\mathbb E(\epsilon_i\epsilon_j^{\ell})|\le c\rho^{d(i,j)}$ for $\ell\in\{1,2\}$.
		 \item [(b)] $|{\rm Cov}(\epsilon_i\epsilon_j,\epsilon_k\epsilon_l)|\le c\rho^{d(\{i,j\},\{k,l\})}$, where for any index set $U,V\subset \mathbb N$ , $d(U,V)=\min_{i\in U, j\in V}d(i,j)$.
		 \item [(c)] For any $h:\mathbb R\mapsto \mathbb R$ uniformly bounded with uniformly bounded derivative, any constant $t$ and any index sets $U$, 
			$|{\rm Cov}(\epsilon_i,h(t S_U))|\le c(1+t|U|)\rho^{d(i,U)}$, where $S_U = \sum_{i \in U} \epsilon_i$.
		 \end{enumerate}		 
\end{assumption}
The correlation decay conditions listed here are less stringent than similar ``weak dependence'' conditions based on pairwise distances \citep{shashkin2010berry,kojevnikov2021limit}.  The exponential decay of correlations resembles the exponential $\rho$-mixing condition in time-series literature \citep{bradley2005basic}. After the statement of our main normal approximation result, we present an example in network auto-regressive model, where \Cref{ass:decay} is theoretically verifiable.  Again, the constant $c$ can be relaxed to grow slowly with $n$.

Our final assumption is about the network neighborhood growth. Intuitively, the network cannot be too densely connected so that the overall dependence is not too strong. 

\begin{assumption}[Network growth]\label{ass:neighbor}
There exist constants $c_1,c>0$ such that
			\begin{equation}\label{eq:neighbor_growth_v3}
	\max\left\{\left(\frac{1}{n}\sum_i N_m^3(i)\right)^{1/2}\,,~~\frac{1}{n}\sum_i N_m^2(i)\right\}\le c e^{c_1m}\, ,
			\end{equation}	
\end{assumption}
where  $N_m(i)$ is the size of the $m$-step neighborhood set $\{j: d(i,j)\le m\}$.
The neighborhood growth condition is already interpretable in the current form.  But it can be further simplified in special cases.
  Let $N_m=n^{-1}\sum_i N_m(i)$ be the average neighborhood size. Eq. \eqref{eq:neighbor_growth_v3} can hold if $N_m$ grows exponentially in $m$ with  a small constant $c_1$ (or slower), while the $m$-step degree sequence $\vec N_m=(N_m(i):1\le i\le n)$ is not too heavy-tailed for $m\asymp \log n$.  Our theory allows for exponentially growing neighborhood size, which is much faster than Euclidean embeddable graphs \citep{weber2020neighborhood}.

\begin{theorem}\label{CLTmixing-reg}
If the sequence $(X,\epsilon,A)$ satisfies \Cref{ass:bound,ass:decay,ass:neighbor} for constants $(c_1,\rho)$ such that
$2c_1 <\log(\rho^{-1})$, then following holds for the least squares estimate $\hat\beta$. 
\begin{enumerate}[label=(\alph*)]
 \item 
      $
     (X^T\Sigma X)^{-1/2} (X^TX) (\hat \beta-\beta)\rightsquigarrow N(0,I)\,.
      $		
			  \item There exists a positive constant $C$, such that for $m=C\log n$ we have
				$$
					      \widehat{X^T\Sigma X}^{-1/2} (X^TX) (\hat \beta-\beta)\rightsquigarrow N(0,I)\,,
					      $$
	      	      where \begin{equation}\label{sigma_est}
				\widehat{X^T\Sigma X}_{rs}=\sum_{i=1}^n\sum_{j:d(i,j)\le m} x_{ir}x_{js}\hat\epsilon_i \hat\epsilon_j.
	      \end{equation}
	     with $\hat\epsilon_i=Y_i-x_{i\cdot}^T\hat\beta$ being the fitted residual at sample point $i$.
			\end{enumerate}	
		\end{theorem}

	
  Now assume that the noise vector $\epsilon$ follows an network autoregressive (NAR) model
	 $$\epsilon_i=\sum_{j}w_{ij} \epsilon_j+U_i\,,~~1\le i\le n\,,$$
	 where $(U_i:i\ge 1)$ is a sequence of independent centered random variables with uniformly bounded $\ell_4$ norm.
	\begin{proposition}\label{pro:NAR_condition}
		If the noise vector $\epsilon$ satisfies the NAR model such that
			\begin{enumerate}
		\item [(i)] \begin{equation}\label{eq:cond_nar}
		\sup_i\sum_{j}|w_{ij}|\equiv \tau < 1\,;
	\end{equation}
	and
	\item [(ii)] $A_{ij}=0$ implies $w_{ij}=0$ for all $1\le i,j\le n$,
	\end{enumerate}
then it satisfies \Cref{ass:decay} with $\rho=\tau^{1/2}$.
	\end{proposition}

\section{Data-driven estimate for the variance components}
Now we turn to inference problems related to the regression coefficients $\beta$.  According to \Cref{CLTmixing-reg}, the asymptotic variance of the OLS estimate $\hat\beta$ is $n(X^TX)^{-1}(X^T\Sigma X)(X^T X)^{-1}$. Since $X$ is observed, we only need to estimate the $p\times p$ matrix $X^T \Sigma X$, which can be achieved by a plug-in formula \eqref{sigma_est}.  Here we explain the idea behind the formula \eqref{sigma_est}.  For each $(r,s)\in\{1,...,p\}^2$, the $(r,s)$-entry of $X^T\Sigma X$ is
$$
\mathbb E\sum_{1\le i,j\le n} x_{ir}x_{js}\epsilon_i\epsilon_j\,.
$$
The correlation decay condition in \Cref{ass:decay}(a) implies that the contribution from the $(i,j)$-pair in the above sum will be negligible when $d(i,j)$ is sufficiently large.  Therefore, we may achieve a better bias-variance trade-off if we truncate the summation based on the distance between nodes $i$ and $j$.  Here the neighborhood range $m$ is a tuning parameter of the estimation procedure and needs to be chosen in a data-driven manner. Interestingly, the optimal choice of $m$ indeed depends on the particular inference we want to make about $\beta$, as we explain next.


Suppose we want to test $H_0: a^T \beta=0$ for some $a\in\mathbb R^p$.  Then the CLT result in \Cref{CLTmixing-reg} implies that, under suitable conditions,
\begin{equation}\label{eq:var_a}
\sqrt{n}a^T(\hat\beta-\beta)\stackrel{d}{\approx} N(0, \sigma_a^2)
\end{equation}
with
$$
\sigma_a^2= n a^T(X^TX)^{-1}(X^T\Sigma X)(X^TX)^{-1}a\,.
$$
Therefore, the network dependence may have different interaction with the design matrix $X$ and contrast vector $a$. To illustrate this interaction, consider a simple case where
the design matrix $X=(x_{\cdot 1},~x_{\cdot 2},~x_{\cdot 3})$ consists of three columns: 
\begin{enumerate}
\item the first column is an intercept column $x_{\cdot 1}=(1,...,1)^T$;
\item the second column $x_{\cdot 2}\sim N(0, \Sigma)$, independent of everything else;
\item the third column $x_{\cdot 3}\sim N(0, I_n)$, independent of everything else.
\end{enumerate}
Consider three different choices of vector $a$: $a_1=(1,0,0)^T$, $a_2=(0,1,0)^T$, $a_3=(0,0,1)^T$.



Assume in addition that the diagonal entries of $\Sigma$ are all $1$.  If we replace $X^TX/n$ by its expected value $I_3$, then for each $a_k$ ($k=1,2,3$) the asymptotic variance becomes
$$
\sqrt{n} a_k^T(\hat\beta-\beta)\stackrel{d}{\approx} N(0, x_{\cdot k}^T\Sigma x_{\cdot k}/n)
$$
where
$x_{\cdot k}$ is the $k$th column of $X$. We find that 
\begin{itemize}
	\item when $k=1$, the asymptotic variance equals the entry-wise sum of $\Sigma$, hence the off-diagonal entries of $\Sigma$ matter when their sum is large.  
	\item when $k=2$, the asymptotic variance approximately equals ${\rm tr}(\Sigma^2)$, hence the off-diagonal entries of $\Sigma$ matter when their squared sum is large.
	\item when $k=3$, the asymptotic variance approximately equals ${\rm tr}(\Sigma)$, and the off-diagonal entries of $\Sigma$ do not matter.
\end{itemize}
The approximations claimed for $k=2,3$ in the above statements follow from the Hansen-Wright inequality \citep{hanson1971bound,rudelson2013hanson}.  This example illustrates that the dependence between entries of $\epsilon$ may or may not bias the inference, depending on how the covariance matrix $\Sigma$ is aligned with the matrix $X(X^TX)^{-1}aa^T(X^TX)^{-1}X^T$.
Thus we recommend choosing $m$ according to the vector $a$.



Let $\gamma_a\in \mathbb R^n = \sqrt{n}X(X^TX)^{-1}a$. Then the asymptotic variance of $a^T(\hat\beta-\beta)$ is
$\sigma_a^2=\gamma_a^T\Sigma\gamma_a$.  The truncated plug-in estimate at radius $m$ is
$$
\hat\sigma_a^2(m) = \sum_{i}\sum_{j:d(i,j)\le m} \gamma_{a,i}\gamma_{a,j}\hat\epsilon_i\hat\epsilon_j\,.
$$
The question reduces to choosing $m$ just enough to achieve a good trade-off between the bias and variance of $\hat \sigma_a^2(m)$.  Define $\Delta_a(m)\equiv \hat\sigma^2_a(m) - \hat\sigma^2_a(0)$. To calibrate the variability caused by including off-diagonal entries of $\Sigma$, we adapt the idea of Moran's I \citep{moran1950notes,lee2021network}, which aims at gauging  whether the difference between $\hat\sigma_a^2(m)$ and $\hat\sigma^2_a(0)$ is due to random fluctuation.  Let $\hat\epsilon_1^{*},...,\hat\epsilon_n^{*}$ be a permuted version of the estimated residuals $\hat\epsilon_1,...,\hat\epsilon_n$. Define $\hat \sigma_a^{2*}(m)$ for the permuted residuals accordingly. After permutation, the adjacency matrix $A$ no longer captures the correct dependence structure between the residuals, so the contribution from the off-diagonal terms $\Delta_a^*(m)\equiv \hat \sigma_a^{2*}(m) - \hat \sigma_a^{2*}(0)$ would be mostly from random fluctuation.

We therefore choose a task-specific $m$ as follows. Let $T$ be the number of permutations, and for $1\leq t \leq T$, $\Delta_a^{*t}(m)$ is the off-diagonal contributions calculated for the $t$-th permuted data. We choose $m$ by
\begin{equation}\label{mselection}
\hat m_a = \min\left\{m\ge 0: \frac{1}{T}\sum_{t=1}^T\mathds{1}\left(|\Delta_a(m+1)|\ge | \Delta_{a}^{*t}(m+1)|\right)\le1-\alpha\right\},\end{equation}
where
$1-\alpha$ is the confidence level. We used $\alpha=0.05$ in all numerical studies and the data example. In the case that we have a categorical variable and $a$ is a multiple column contrast matrix, we shall use the operator norm of $\Delta_{a}(m)$ to replace the absolute value in \eqref{mselection}. 


\section{Simulation}\label{sim}

In this section, we demonstrate the finite sample performance of the method, especially in reducing the inflated Type I error that may arise if the underlying network structure is not appropriately taken into account.

For a given network with adjacency matrix $A$, and it's degree-normalized version $A_n$ obtained by dividing each entry of $A$ by its row sum, the dependence among error terms is generated through a matrix $B$, i.e., $\epsilon=B\epsilon_w$ for an $n$-dimensional white noise vector $\epsilon_w$. We use Gaussian white noise in this simulation. We generate $B$ based on auto-regressive models (AR), moving average models (MA), and direct transmission models (DT). For the AR model, $B=(I-\rho A_n)^{-1}$. For the MA model, $B=I+\rho A_n$, and in the case of DT model, $B=(I+\rho A_n)^2$. The DT model extends the MA model by considering the influence of longer ranges, which is a modification of the setting explored in \cite{lee2021network}. The parameter $\rho$ controls the magnitude of dependence, which is set to either $0.2$ or $0.4$.  Three columns of predictor variables are considered; $x_{\cdot 1}=1$ (intercept), $x_{\cdot 2}=B\epsilon_{x_2}$ and $x_{\cdot 3}=\epsilon_{x_3}$, where $\epsilon_{x_2}$ and $\epsilon_{x_3}$ are independently generated standard normal vectors of length $n$. Without loss of generality in assessing Type I error rate, the true $\beta$ coefficients are set to zero, and we report the empirical Type I error under the $\alpha = 0.05$ nominal level.

 We consider two types of network structure. In Simulation I, the network is generated from a 4-block Stochastic Block Model with a sample size of $n=300$ and an equal proportion of nodes in each block. The connection probabilities within and between these groups are represented by a symmetric matrix $\gamma P\in [0,1]^{4\times 4}$. The parameter $\gamma$ adjusts the overall density of the network and is set to $0.5, 1,$ or $1.5$.  The off-diagonal values of $P$  is set to be 0.005 and the diagonal values are set to be (0.005, 0.010, 0.015, 0.020). Nodes without any connections are removed afterwards. Once $A$ is generated, it remains fixed throughout the simulation repetitions. 

In Simulation II, we utilize the network data and the seven covariates from our student data example, the details of which are presented in the next section. The sample size is 244 and all nodes are initially connected to at least one other node.

\begin{table}[!ht] 
    \centering
    \footnotesize
    \begin{tabular}{|c|c|c|c|c|c|c|c|c|c|c|c|c|}
    \hline
       & \multirow{2}{*}{$\rho$}&  \multirow{2}{*}{$m$} & \multicolumn{3}{|c|}{$\gamma$=0.5}&  \multicolumn{3}{|c|}{$\gamma$=1} &  \multicolumn{3}{|c|}{$\gamma$=1.5}    \\  \cline{4-12}  
        &&& $x_{\cdot 1}$  &$x_{\cdot 2}$ & $x_{\cdot 3}$ & $x_{\cdot 1}$ & $x_{\cdot 2}$ & $x_{\cdot 3}$ & $x_{\cdot 1}$ & $x_{\cdot 2}$ & $x_{\cdot 3}$    \\ \hline  
      \multirow{6}{*}{AR}  &     \multirow{3}{*}{0.2}&$m=0$ &0.11 & 0.06  & 0.06  & 0.11  & 0.06  & 0.06  & 0.11  & 0.07  & 0.05  \\
       && $\hat m_a$ & 0.07 & 0.06  & 0.06  & 0.07  & 0.05  & 0.06  & 0.06  & 0.07  & 0.05  \\
       \cline{2-12}
         &     \multirow{3}{*}{0.4}&$m=0$ &0.16 & 0.11  & 0.06  & 0.21  & 0.1 & 0.06  & 0.21  & 0.08  & 0.06  \\
       && $\hat m_a$ & 0.05 & 0.08  & 0.06  & 0.07  & 0.07  & 0.06  & 0.08  & 0.07  & 0.06  \\
       \hline  
       
      \multirow{6}{*}{MA}  &     \multirow{3}{*}{0.2}&$m=0$ &0.11 & 0.07  & 0.06  & 0.12  & 0.07  & 0.07  & 0.12  & 0.06  & 0.05  \\       
       && $\hat m_a$ & 0.05 & 0.07  & 0.06  & 0.06  & 0.07  & 0.07  & 0.06  & 0.06  & 0.05  \\
       \cline{2-12}
         &     \multirow{3}{*}{0.4}&$m=0$ &0.15 & 0.09  & 0.06  & 0.17  & 0.07  & 0.07  & 0.17  & 0.06  & 0.07  \\      
       && $\hat m_a$ &0.07  & 0.06  & 0.06  & 0.06  & 0.05  & 0.06  & 0.06  & 0.05  & 0.07  \\
       \hline  
       
         \multirow{6}{*}{DT}  &     \multirow{3}{*}{0.2}&$m=0$ &0.15  & 0.09  & 0.06  & 0.15  & 0.09  & 0.06  & 0.17  & 0.09  & 0.07  \\       
       && $\hat m_a$ & 0.07 & 0.07  & 0.06  & 0.06  & 0.08  & 0.06  & 0.07  & 0.08  & 0.07  \\
       \cline{2-12}
         &     \multirow{3}{*}{0.4}&$m=0$ &0.22 & 0.16  & 0.04  & 0.25  & 0.14  & 0.06  & 0.26  & 0.13  & 0.07  \\      
       && $\hat m_a$ &0.07  & 0.07  & 0.04  & 0.08  & 0.05  & 0.06  & 0.09  & 0.06  & 0.07  \\
       \hline  

            \end{tabular} 
            \caption{Type I error of $\beta$ coefficients in Simulation I.}
            \label{sbm}
\end{table}

\begin{table}[!ht]
    \centering
    \footnotesize
    \begin{tabular}{|c|c|c|c|c|c|c|c|c|c|c|}
    \hline
       &$\rho$& m & Intercept  &hh.size & age  & sex & play & homework  & bio.parents  \\ \hline  
      \multirow{6}{*}{AR}  &     \multirow{3}{*}{0.2}&m=0 & 0.05  & 0.06  & 0.05  & 0.1 & 0.06  & 0.05  & 0.07\\ 
       && $\hat m_a$ & 0.05 & 0.06  & 0.05  & 0.08  & 0.06  & 0.05  & 0.07\\ 
       \cline{2-10}
         &     \multirow{3}{*}{0.4}&m=0 &0.06 & 0.06  & 0.06  & 0.2 & 0.06  & 0.06  & 0.09\\ 
       && $\hat m_a$& 0.06  & 0.06  & 0.05  & 0.09  & 0.06  & 0.05  & 0.09\\ 
         \hline
            \multirow{6}{*}{MA}  &     \multirow{3}{*}{0.2}&m=0 & 0.07  & 0.06  & 0.06  & 0.08  & 0.06  & 0.05  & 0.08\\ 
       && $\hat m_a$ & 0.06 & 0.06  & 0.06  & 0.07  & 0.06  & 0.05  & 0.09\\ 
       \cline{2-10}
         &     \multirow{3}{*}{0.4}&m=0 &0.06 & 0.06  & 0.05  & 0.16  & 0.05  & 0.06  & 0.09\\ 
       && $\hat m_a $ & 0.06  & 0.06  & 0.05  & 0.07  & 0.05  & 0.06  & 0.09\\ 
         \hline
           \multirow{6}{*}{DT}  &     \multirow{3}{*}{0.2}&m=0 &0.05  & 0.06  & 0.04  & 0.15  & 0.04  & 0.04  & 0.09\\ 
       && $\hat m_a$ &0.05  & 0.06  & 0.04  & 0.07  & 0.04  & 0.04  & 0.09\\ 
       \cline{2-10}
         &     \multirow{3}{*}{0.4}&m=0 &0.07 & 0.09  & 0.06  & 0.24  & 0.05  & 0.06  & 0.08\\ 
       && $\hat m_a$ &0.06  & 0.08  & 0.06  & 0.09  & 0.05  & 0.06  & 0.08\\ 
         \hline

            \end{tabular}
            \caption{Type I error of $\beta$ coefficients for Simulation II.}
            \label{realsim}
\end{table}

The results of the first simulation are presented in Table \ref{sbm}. The table consists of nine columns representing the outcomes for testing $\beta_j=0$ for three predictive variables ($j=1,2,3$) under three values of the network density parameter ($\gamma$). The rows correspond to the results obtained from three different models (AR, MA, DT) and two different values of $\rho$. We examine type I errors at nominal level $0.05$ under two schemes: no adjustment ($m=0$) and the task-specific selection as described in Section 3 ($\hat m_a$). The maximum $m$ in the search range is set to be 6. Neglecting the potential dependency due to networks, the type I errors for $x_{\cdot 1}$ and $x_{\cdot 2}$ are found to be inflated across nearly all of the settings. The inflation is particularly severe when the network dependence is strong ($\rho=0.4$) or the network dependency is long range (DT model). The type I errors for $x_{\cdot 3}$ remain close to the nominal level. 
This result is consistent with the phenomenon described in Section 3. The dependence between entries of $\epsilon$ may or may not bias the inference, depending on how network dependency is aligned with the matrix $X(X^TX)^{-1}aa^T(X^TX)^{-1}X^T$, where $X$ is the design matrix and $a$ is the contrast vector corresponding to the hypothesis of interest. 
In cases where the Type I error is inflated, applying adjustments using the recommended selection scheme proves effective in reducing the type I error. The findings are similar across different levels of network density. 

The results for Simulation II are presented in Table \ref{realsim}. The table displays seven columns representing the Type I errors of testing the coefficients for each of the seven  predictor variables. Among the variables examined, the ``sex" variable consistently exhibits inflated Type I errors across different settings. As explained in the data analysis section, the ``sex" variable has an obvious alignment with the network structure and therefore the design column is likely to be aligned with the dependence structure. Adjusting $m$ is shown to be helpful in reducing the Type I errors.

\section{Student Data Example} \label{student}

To further illustrate the method, we apply the proposed method to a student data set with a known network structure. The data consist of survey responses collected in 2018 from three public primary schools in Tanzania \citep{schurz2020distributional}. These responses include students' school grade scores and their background characteristics. Additionally, students were asked to provide information on up to three of their closest friends, including their interaction patterns such as how often they play or do homework together.

After removing observations with missing data, the size of the three schools are as follows: 244, 161, and 225. We primarily focus on the largest school, while considering the other schools as additional data sets to enhance the interpretation of our findings. Table \ref{var} provides a summary of the data. We treat the ``sex" and ``bio.parents" variables as categorical variables.  Two students (nodes) are considered to have an edge in the network if at least one of them reported the other as a close friend.  
We note that the data also contain religious information and the majority are Muslim. Controlling this variable in the regression model would slightly alter the fitted coefficients of other variables but not change any of the significant/non-significant conclusions or the signs of the fitted coefficients. For the purpose of illustrating the proposed method, all results and discussions for this data analysis and Simulation II remain the same regardless of whether the religious variable is controlled or not. The data also contain a variable indicating the number of children in the household, which is not included in the regression due to its high collinearity with the household size variable.

\begin{table}[!ht]
    \centering
    \footnotesize
    \begin{tabular}{|l|l|l|l|l|}
    \hline
        Variable &  Mean/frequency & sd & Definition \\ \hline

        School grade score ($Y$)&  451.5  & 112.4 &\\ \hline
        Household size & 5.27 & 2.2 & Number of people living in the household\\ \hline
        Age & 12.6 & 1.0& \\ \hline
        Sex &   &  &  \\
          1: female &  0.52&&\\ 
          0: male   & 0.48&&\\ \hline
         
        How often play with friends &  2.35 & 0.9 & {0= never, 1=once a week, 2=two or three  }\\ \cline{1-3}
        How often do homework with friends &  1.47 & 1.2 &times a week, 3=everyday \\ \hline
        Living with biological parents & & &\\
        0: both &  0.47&&\\ 
        1 : only mother / only father & 0.44&&\\
        2 : none & 0.09&& \\ \hline
        In-network friends  &  4.5 & 1.7 & Degree of nodes in friendship network \\ \hline
      \end{tabular} 
           \caption{A summary of the variables and the network characteristics. }
            \label{var}

\end{table}

\begin{figure} [!ht]
\vspace{0.5in}
\includegraphics[width=0.5\textwidth]{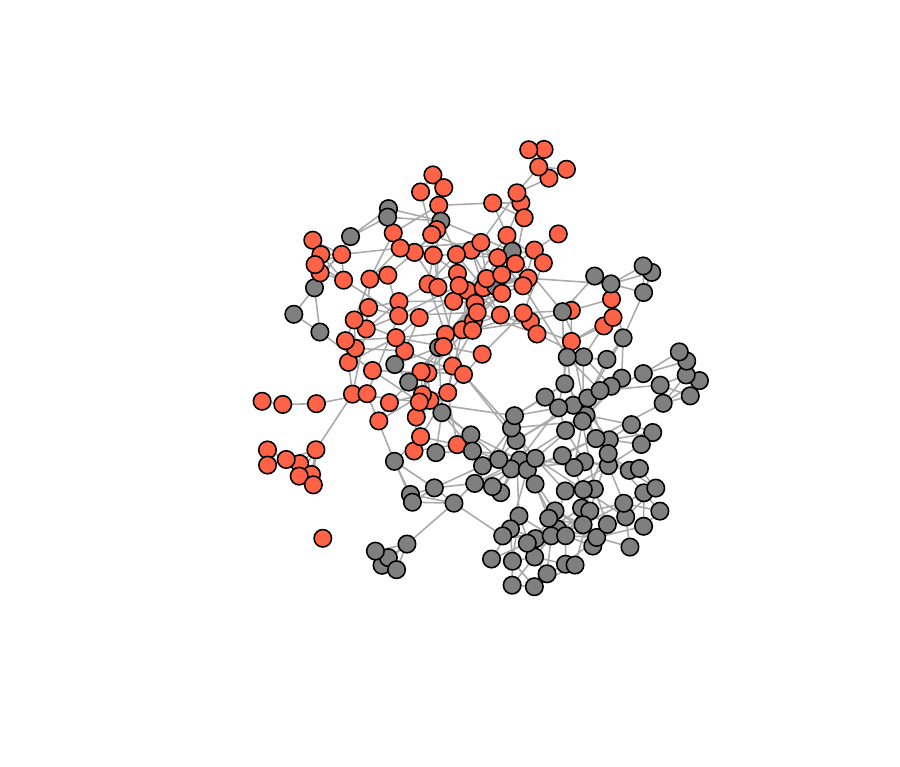}
\includegraphics[width=0.4\textwidth]{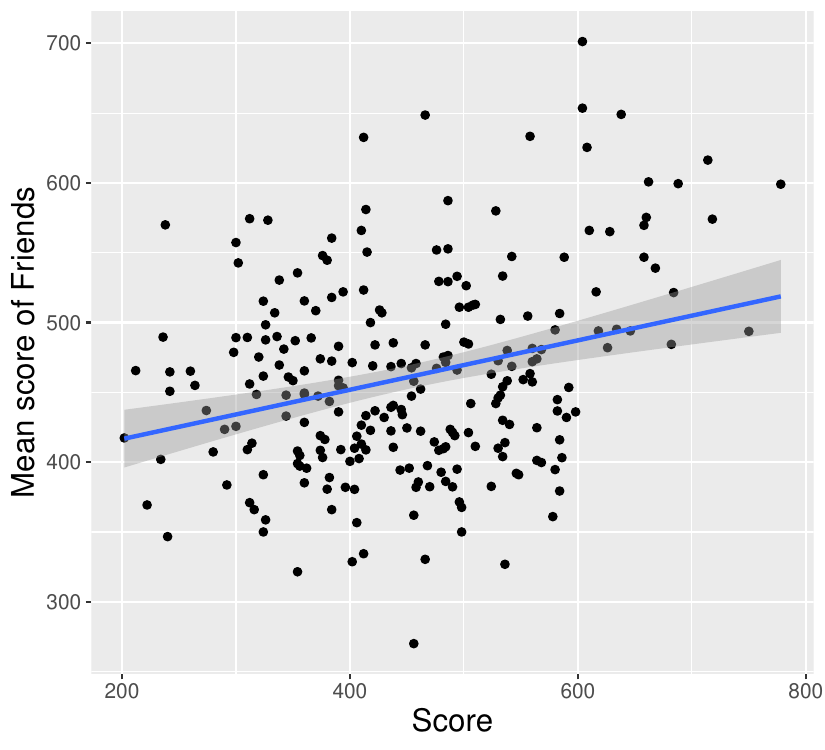}
\caption{ Left: Network structure with females represented in gray and males in orange. Right: Scatter plot between the school grade score of students and the average score of their friends (Pearson's correlation coefficient =0.28.)  }
\label{student}
\end{figure}

We consider the student's school grade score as the response variable and include other covariates as predictor variables. The authors of the paper \cite{schurz2020distributional}, where the data set was first released, discovered a positive correlation between the school grade score of students and those of their friends. Additionally, the authors emphasized the significant presence of gender-based segregation within these social networks; see Figure \ref{student}. These findings collectively suggest the need of dealing with the network dependency, especially for the inference of the ``sex'' variable. We therefore apply the proposed method to adjust the network dependency in the inference of regression coefficients.


\begin{table}[!ht]
\footnotesize
    \centering
    \begin{tabular}{|c|c|c|c|}
    \hline
             & $m=0$  & $\hat m_a$  \\ \hline

        (Intercept) & 0.00* & 0.00*   \\ \hline
        hh.size & 0.09 & 0.35   \\ \hline
        age & 0.19  & 0.19  \\ \hline
        sex & 0.02* & 0.16   \\ \hline
        play & 0.62 & 0.62  \\ \hline
        homework & 0.03*  & 0.03*  \\ \hline
        bio.parents & 0.21 & 0.21  \\ \hline
    \end{tabular} 
    \caption{P-values of each covariate with different $m$ selection for the largest school.}
     \label{pval}
\end{table}

Table \ref{pval} shows the p-values for each covariate with and without network adjustment. When no adjustment is applied, the variables ``homework" and ``sex" exhibit statistical significance at the 0.05 level. However, after network adjustment, the ``sex" variable is not statistically significant anymore.

\begin{table}[!ht]
\footnotesize
    \centering
    \begin{tabular}{|c|c|c|c|}
    \hline
            & main  school & test school 1 & test school 2 \\ \hline

        (Intercept) & 670.4 & 581.8 & 363.1  \\ \hline
        hh.size & -5.5 & 3.8 & 7.2   \\ \hline
        age & -10.0 & -12.7 & 10.9  \\ \hline
        sex (female) & -33.7 & 7.2 & 10.2   \\ \hline
        play & -4.0 & -2.8 & -5.7  \\ \hline
        homework & -13.6 & -2.6  & -4.0  \\ \hline
        bio.parents1 (only father/mother) & -27.3 & 0.55 & -40.8  \\ \hline
        bio.parents2 (none) & -31.5 & -24.5 & -48.0 \\ \hline
    \end{tabular}
        \caption{Regression coefficients for three schools.}
     \label{other}
   \end{table}

In order to gain a better understanding of this result, we conduct separate linear regression analyses for each of the other two schools, which have sizes of 161 and 225, respectively. The regression coefficients are summarized in Table \ref{other}. We observe that the regression coefficients of the ``homework" variable have the same sign across all three schools. However, the signs of the regression coefficient of the ``sex" variable are positive in the main school and negative in the other schools. Recall that in Simulation II the ``sex" variable had an inflated Type I error rate when data were generated from a model without the ``sex" predictor. These observations indicate that the significant result for the ``sex" variable using regular OLS method ($m =0$) could be a spurious finding due to the network dependency.


\section{Discussion}\label{sec:disc}
As long as the decay of correlations is exponential as the distance increases, we show that the ordinary least squares estimate remains consistent and asymptotically normal even when the neighborhood sizes grow exponentially. While this result extends the applicability of the popular OLS-based inference under network dependency, it calls for further investigation and innovation on adaptive inference, especially in estimating the asymptotic variance.  In practice, the network is given but the decay speed of the correlations is unknown, making the choice of truncation radius $m$ a challenging task.  An adaptive choice of bias-variance trade-off, contingent upon an unknown regularity parameter, resembles the model selection problem in non-parametric and high-dimensional inferences akin to methods like Lepskii's method and cross-validation \citep[see][and references therein]{heidenreich2013bandwidth,wu2020survey}.  It would be interesting to further develop and study these methods under the network dependence context.

The Gaussian comparison results for general dependent data beyond Euclidean embeddable cases are another useful probability tools for statistical inference, and can potentially be extended in several directions, including generalized linear models \citep{bonney1987logistic,daskalakis2019regression}, semiparametric models \citep{bravo2022misspecified}, and high-dimensional data \citep{kurisu2023gaussian,chang2024central}.

\section*{Acknowledgements}
Jing Lei's research is partially supported by NSF grants DMS-2015492 and DMS-2310764. Kehui Chen's research is partially supported by NSF grant DMS-2210402.

\bibliographystyle{apa-good}
\bibliography{networkcombined,data}

\begin{thebibliography}{24}
\expandafter\ifx\csname natexlab\endcsname\relax\def\natexlab#1{#1}\fi
\expandafter\ifx\csname url\endcsname\relax
  \def\url#1{{\tt #1}}\fi
\expandafter\ifx\csname urlprefix\endcsname\relax\def\urlprefix{URL }\fi

\bibitem[{Baldi \& Rinott(1989)}]{baldi1989normal}
Baldi, P., \& Rinott, Y. (1989).
\newblock On normal approximations of distributions in terms of dependency
  graphs.
\newblock {\em The Annals of Probability\/}, (pp. 1646--1650).

\bibitem[{Bonney(1987)}]{bonney1987logistic}
Bonney, G.~E. (1987).
\newblock Logistic regression for dependent binary observations.
\newblock {\em Biometrics\/}, (pp. 951--973).

\bibitem[{Bradley(2005)}]{bradley2005basic}
Bradley, R.~C. (2005).
\newblock Basic properties of strong mixing conditions. a survey and some open
  questions.
\newblock {\em Probability Surveys\/}, {\em 2\/}, 107--144.

\bibitem[{Bravo(2022)}]{bravo2022misspecified}
Bravo, F. (2022).
\newblock Misspecified semiparametric model selection with weakly dependent
  observations.
\newblock {\em Journal of Time Series Analysis\/}, {\em 43\/}(4), 558--586.

\bibitem[{Chang et~al.(2024)Chang, Chen, \& Wu}]{chang2024central}
Chang, J., Chen, X., \& Wu, M. (2024).
\newblock Central limit theorems for high dimensional dependent data.
\newblock {\em Bernoulli\/}, {\em 30\/}(1), 712--742.

\bibitem[{Chen \& Shao(2004)}]{chen2004normal}
Chen, L.~H., \& Shao, Q.-M. (2004).
\newblock Normal approximation under local dependence.
\newblock {\em The Annals of Probability\/}, {\em 32\/}(3), 1985--2028.

\bibitem[{Daskalakis et~al.(2019)Daskalakis, Dikkala, \&
  Panageas}]{daskalakis2019regression}
Daskalakis, C., Dikkala, N., \& Panageas, I. (2019).
\newblock Regression from dependent observations.
\newblock In {\em Proceedings of the 51st Annual ACM SIGACT Symposium on Theory
  of Computing\/}, (pp. 881--889).

\bibitem[{Gaetan \& Guyon(2010)}]{gaetan2010spatial}
Gaetan, C., \& Guyon, X. (2010).
\newblock {\em Spatial Statistics and Modeling\/}, vol.~90.
\newblock Springer.

\bibitem[{Hanson \& Wright(1971)}]{hanson1971bound}
Hanson, D.~L., \& Wright, F.~T. (1971).
\newblock A bound on tail probabilities for quadratic forms in independent
  random variables.
\newblock {\em The Annals of Mathematical Statistics\/}, {\em 42\/}(3),
  1079--1083.

\bibitem[{Heidenreich et~al.(2013)Heidenreich, Schindler, \&
  Sperlich}]{heidenreich2013bandwidth}
Heidenreich, N.-B., Schindler, A., \& Sperlich, S. (2013).
\newblock Bandwidth selection for kernel density estimation: a review of fully
  automatic selectors.
\newblock {\em AStA Advances in Statistical Analysis\/}, {\em 97\/}, 403--433.

\bibitem[{Huber(1967)}]{huber1967under}
Huber, P.~J. (1967).
\newblock Under nonstandard conditions.
\newblock In {\em Proceedings of the Fifth Berkeley Symposium on Mathematical
  Statistics and Probability: Weather modification\/}, vol.~5, (p. 221). Univ
  of California Press.

\bibitem[{Kojevnikov et~al.(2021)Kojevnikov, Marmer, \&
  Song}]{kojevnikov2021limit}
Kojevnikov, D., Marmer, V., \& Song, K. (2021).
\newblock Limit theorems for network dependent random variables.
\newblock {\em Journal of Econometrics\/}, {\em 222\/}(2), 882--908.

\bibitem[{Kurisu et~al.(2023)Kurisu, Kato, \& Shao}]{kurisu2023gaussian}
Kurisu, D., Kato, K., \& Shao, X. (2023).
\newblock Gaussian approximation and spatially dependent wild bootstrap for
  high-dimensional spatial data.
\newblock {\em Journal of the American Statistical Association\/}, (pp. 1--13).

\bibitem[{Lee \& Ogburn(2021)}]{lee2021network}
Lee, Y., \& Ogburn, E.~L. (2021).
\newblock Network dependence can lead to spurious associations and invalid
  inference.
\newblock {\em Journal of the American Statistical Association\/}, {\em
  116\/}(535), 1060--1074.

\bibitem[{Liang \& Zeger(1986)}]{liang1986longitudinal}
Liang, K.-Y., \& Zeger, S.~L. (1986).
\newblock Longitudinal data analysis using generalized linear models.
\newblock {\em Biometrika\/}, {\em 73\/}(1), 13--22.

\bibitem[{Moran(1950)}]{moran1950notes}
Moran, P.~A. (1950).
\newblock Notes on continuous stochastic phenomena.
\newblock {\em Biometrika\/}, {\em 37\/}(1/2), 17--23.

\bibitem[{Ogburn et~al.(2017)Ogburn, Sofrygin, van~der Laan, \&
  Diaz}]{ogburn2017causal}
Ogburn, E.~L., Sofrygin, O., van~der Laan, M., \& Diaz, I. (2017).
\newblock Causal inference for social network data with contagion.
\newblock {\em ArXiv e-prints\/}.

\bibitem[{P{\'e}rez-Escudero et~al.(2014)P{\'e}rez-Escudero, Vicente-Page,
  Hinz, Arganda, \& De~Polavieja}]{perez2014idtracker}
P{\'e}rez-Escudero, A., Vicente-Page, J., Hinz, R.~C., Arganda, S., \&
  De~Polavieja, G.~G. (2014).
\newblock idtracker: tracking individuals in a group by automatic
  identification of unmarked animals.
\newblock {\em Nature methods\/}, {\em 11\/}(7), 743--748.

\bibitem[{Rudelson \& Vershynin(2013)}]{rudelson2013hanson}
Rudelson, M., \& Vershynin, R. (2013).
\newblock Hanson-wright inequality and sub-gaussian concentration.
\newblock {\em Electronic Communications In Probability\/}, {\em 18\/}, 1--9.

\bibitem[{Sch{\"u}rz et~al.(2020)Sch{\"u}rz, Alem, Kocher, Carlsson, \&
  Lindahl}]{schurz2020distributional}
Sch{\"u}rz, S., Alem, Y., Kocher, M., Carlsson, F., \& Lindahl, M. (2020).
\newblock Distributional preferences in adolescent peer networks.
\newblock Tech. rep., IHS Working Paper.

\bibitem[{Shashkin(2010)}]{shashkin2010berry}
Shashkin, A. (2010).
\newblock A berry--esseen type estimate for dependent systems on transitive
  graphs.
\newblock In {\em Advances in Data Analysis\/}, (pp. 151--156). Springer.

\bibitem[{Weber(2020)}]{weber2020neighborhood}
Weber, M. (2020).
\newblock Neighborhood growth determines geometric priors for relational
  representation learning.
\newblock In {\em International Conference on Artificial Intelligence and
  Statistics\/}, (pp. 266--276). PMLR.

\bibitem[{White(1980)}]{white1980heteroskedasticity}
White, H. (1980).
\newblock A heteroskedasticity-consistent covariance matrix estimator and a
  direct test for heteroskedasticity.
\newblock {\em Econometrica: journal of the Econometric Society\/}, (pp.
  817--838).

\bibitem[{Wu \& Wang(2020)}]{wu2020survey}
Wu, Y., \& Wang, L. (2020).
\newblock A survey of tuning parameter selection for high-dimensional
  regression.
\newblock {\em Annual review of statistics and its application\/}, {\em 7\/},
  209--226.

\end{thebibliography}

\appendix
 \section{Proof of main results}
\subsection{Proof of \Cref{CLTmixing-reg} part (a):} 

Since $\hat\beta-\beta=(X^TX)^{-1}X^T\epsilon$, it suffices to prove that for any $\alpha\in \mathbb R^p$, $\|\alpha\|_2=1$ we have
$$\alpha^T(X^T\Sigma X)^{-1/2}X^T\epsilon \rightsquigarrow N(0,1)\,.$$

Let $z=\sqrt{n}X(X^T\Sigma X)^{-1/2}\alpha$, then $|z_i|\le c^{3/2}\sqrt{p}=O(1)$ by \Cref{ass:bound},
and we can write, letting $\xi_i=z_i\epsilon_i$
$$
\alpha^T(X^T\Sigma X)^{-1/2}X^T\epsilon = \frac{\sum_{i=1}^n z_i \epsilon_i}{\sqrt{n}}=\frac{\sum_{i=1}^n \xi_i}{\sqrt{n}}
$$

Define $S = \sum_{i}\xi_i$ and $S_{i}(m) =\sum_{j:d(i,j)\le m} \xi_{j}$. Notice that we often omit $m$ in the parenthesis in these notations whenever its clear in the context. Let $\sigma^2 = {\rm Var}(S)\equiv n$.

Let $h\in\mathbb C_b^3(\mathbb R\mapsto\mathbb R)$.  According to Stein's method, we only need to show
$$
\mathbb E\left[ \frac{S}{\sigma}h\left(\frac{S}{\sigma}\right) - h'\left(\frac{S}{\sigma}\right)\right]\rightarrow 0\,.
$$
For any $m$, 
\begin{align*}
	&(S/\sigma) h(S/\sigma)-h'(S/\sigma)\\
	=&\sum_{i} (\xi_i/\sigma)\left[h((S-S_i)/\sigma) + h(S/\sigma)-h((S-S_i)/\sigma)\right] - h'(S/\sigma)\\
	\stackrel{(a)}{=}&\sum_i \frac{\xi_i}{\sigma} h\left(\frac{S-S_i}{\sigma}\right) \\
	&+\sum_i \frac{\xi_i}{\sigma} \frac{S_i}{\sigma} h'\left(\frac{S}{\sigma}\right)\\
	&-\sum_i\frac{\xi_i}{\sigma} \left(\frac{S_i}{\sigma}\right)^2\int_0^1 u h''\left(\frac{S-S_i}{\sigma}+u\frac{S_i}{\sigma}\right)du\\
	&-h'\left(\frac{S}{\sigma}\right)\\
	=&\sum_i \frac{\xi_i}{\sigma} \frac{S_i}{\sigma} h'\left(\frac{S}{\sigma}\right)-h'\left(\frac{S}{\sigma}\right)\\
	&+\sum_i \frac{\xi_i}{\sigma} h\left(\frac{S-S_i}{\sigma}\right) \\
	&-\sum_i\frac{\xi_i}{\sigma} \left(\frac{S_i}{\sigma}\right)^2\int_0^1 u h''\left(\frac{S-S_i}{\sigma}+u\frac{S_i}{\sigma}\right)du
\end{align*}
where step $(a)$ follows from the standard first order expansion of $h(S/\sigma)-h((S-S_i)/\sigma)$.

We need to show the expectation of each term is small.

\paragraph{First term.}

The first term equals
$$
h'\left(\frac{S}{\sigma}\right)\left[\frac{\sum_i \xi_i S_i}{\sigma^2}-1\right]\,,
$$
whose expectation is bounded by, letting $\hat\sigma^2=\sum_{i} \xi_i S_i$,
\begin{align*}
&\|h'\|_\infty \mathbb E \left|\frac{\hat\sigma^2}{\sigma^2}-1\right|\\
\lesssim & n^{-1}\mathbb E|\hat\sigma^2-\sigma^2|\\
\lesssim & n^{-1}\left(\mathbb E|\hat\sigma^2-\sigma_0^2|+|\sigma_0^2+\sigma^2|\right)\,.
\end{align*}
Using the same argument as the variance estimation part, in particular \eqref{eq:var_trunc} and \eqref{eq:tilde_gamma_rs-gamma_rs0-square-bound} we have
\begin{itemize}
	\item $|\sigma^2-\sigma_0^2|\lesssim n^2\rho^m$;
	\item $\mathbb E(\hat\sigma^2-\sigma_0^2)^2\lesssim \sum_i N_m^3(i)+n^2\rho^m \bar N_m^2$\,. 
\end{itemize}

So the total contribution of the first term is at most, recall that $\sigma^2=n$, ignoring constant factors,
\begin{align*}
&n\theta_m+\frac{\left(n^{-1}\sum_i N_m^3(i)\right)^{1/2}}{n^{1/2}}+\theta_m^{1/2}\bar N_m\\
\lesssim & n\rho^m+n^{-1/2}e^{c_1 m}+\rho^{m/2}e^{c_1 m/2}\,, 
\end{align*}
where the inequality follows from \Cref{ass:neighbor}.  Thus the first term is $o(1)$ when
$$
\min\left( \frac{m}{\log n} - \frac{1}{\log(\rho^{-1})}\,,~ \frac{1}{2c_1} - \frac{m}{\log n}\right)
$$
 is positive and bounded away from $0$.

\paragraph{Second term.}

\begin{align*}
	&\mathbb E \xi_i h\left(\frac{S_{-i}}{\sigma}\right)\\
=&{\rm Cov}\left( \xi_i, h\left(\frac{S_{-i}}{\sigma}\right)\right)\\
\le & \|h'\|_\infty \frac{n}{\sigma}\theta_m
\end{align*}
by  \Cref{ass:decay}(c).

The total contribution of the second term is at most
$$
\frac{n^2}{\sigma^2}\theta_m\lesssim n\theta_m = o(1)
$$
whenver $m/ \log n - 1/\log(\rho^{-1})$ is bounded away from $0$.

\paragraph{Third term.}

Since $\|\xi_i\|_3$ is uniformly bounded by a constant for all $i$, we have $\|S_i\|_3\le N_m(i)$.
By H\"{o}lder inequality, the third term is at most
$$
\sum_i\frac{N_m^2(i)}{\sigma^3}= \frac{n^{-1}\sum_{i}N_m^2(i)}{n^{1/2}} \le c n^{-1/2}e^{c_1 m}
$$
which is $o(1)$ if $\frac{1}{2c_1}-\frac{m}{\log n}$ is positive and bounded away from 0.

\subsection{Proof of \Cref{CLTmixing-reg} part (b):}
Now our goal is to establish
      $$
     \widehat{X^T\Sigma X}^{-1/2} (X^TX) (\hat \beta-\beta)\rightsquigarrow N(0,I)\,.
      $$		
for the estimate $\widehat{X^T\Sigma X}^{-1/2}$ given in \eqref{sigma_est} with a suitably chosen $m$. 

We will achieve this by upper bounding the entry-wise estimation error with an $o(n)$ rate. An $o(n)$ error rate is sufficient because by assumption $X^T\Sigma X$ has minimum and maximum eigenvalues at the order $n$. The desired result follows by combining part (a) with Slutsky's theorem.

To simply notation, let $x=x_{\cdot r}$ and $y=x_{\cdot s}$ for $1\le r < s\le p$ be two columns of $X$.
We want to approximate 
$$\Gamma_{rs}=(X^T\Sigma X)_{rs}=\sum_{1\le i,j\le n} x_i y_j \Sigma_{ij}=\mathbb E\sum_{i,j}x_i y_j \epsilon_i\epsilon_j$$


For a given truncation radius $m$, define
$$
\Gamma_{rs,0}=\sum_{i}\sum_{j:d(i,j)\le m} x_i y_j \Sigma_{ij}
$$
We have, by \Cref{ass:decay}(a) which simplies $\Sigma_{ij}\le c \rho^{d(i,j)}$
\begin{align}
	\Gamma_{rs,0}-\Gamma_{rs} = & \sum_{i}\sum_{j:d(i,j)>m}x_iy_j \Sigma_{ij}\nonumber\\
	                          \le & c^3 n^2 \rho^m\label{eq:var_trunc}\\
	                          = & o(n)\,.\nonumber
\end{align}
as long as $n \rho^m =o(1)$ which is satisfied when
$m/\log n - 1/\log(\rho^{-1})$ is positive and bounded away from $0$.

Let $\tilde \Gamma_{rs}=\sum_{i}\sum_{j:d(i,j)\le m} \epsilon_i\epsilon_j x_i y_j$.

Then, using the fact that $\hat\epsilon_i-\epsilon_i=x_{i\cdot}^T(\beta-\hat\beta)$
\begin{align*}
	|\hat\Gamma_{rs}-\tilde\Gamma_{rs}| =& \left|\sum_{i}\sum_{j:d(i,j)\le m}\left[ (\hat\epsilon_i-\epsilon_i)(\hat \epsilon_j-\epsilon_j)+(\hat\epsilon_i-\epsilon_i)\epsilon_j+(\hat\epsilon_j-\epsilon_j)\epsilon_i\right]x_i y_j\right|\\
	\le & \sum_{i} N_m(i)\left(c^4 p\|\hat\beta-\beta\|^2 +2c^3\sqrt{p}\|\hat\beta-\beta\|\right)\\
	\lesssim & n e^{c_1 m /2} \|\hat\beta-\beta\|  = o_P(n)
\end{align*}
whenever $1/c_1-m/\log n$ is positive and bounded away from $0$, because we have already established in part(a) of \Cref{CLTmixing-reg} that $\hat\beta-\beta=O_P(1/\sqrt{n})$.

The last step is to bound $\tilde\Gamma_{rs}-\Gamma_{rs,0}$. For $u=(i,j)\in [n]^2$, let $W_{u}=(\epsilon_i\epsilon_j-\Sigma_{ij})x_i y_j$.  Let $\mathcal U=\{(i,j)\in [n]^2: d(i,j)\le m\}$.
\begin{align}
\mathbb E	\left(\tilde\Gamma_{rs}-\Gamma_{rs,0} \right)^2= & \mathbb E\left(\sum_{u\in\mathcal U}W_u\right)^2\nonumber\\
=&\mathbb E\sum_{u,v\in \mathcal U} W_u W_v\nonumber\\
=& \mathbb E\sum_{u\in \mathcal U} \sum_{v:v\in\mathcal U,~d(u,v)\le m} W_u W_v +\mathbb E\sum_{u\in\mathcal U}\sum_{v:v\in\mathcal U,~d(u,v)>m} W_u W_v\,.\label{eq:tilde_gamma_rs-gamma_rs0-decomp}
\end{align}
In the first term, $\mathbb E W_u W_v\le \|W_u\|_2\|W_v\|_2=O(1)$ by \Cref{ass:bound}.
Each $(u,v)$ pair in this sum corresponds to a quadruplet $(i,j,k,l)$ such that $d(j,i)\le m$, $d(k,j)\le m$, $d(k,l)\le m$. So the first term is bounded by, up to a constant factor,
\begin{align*}
&\sum_i \sum_{j:d(i,j)\le m} \sum_{k: d(k,j)\le m}\sum_{l:d(k,l)\le m} 1\\
= & \sum_i \sum_{j:d(i,j)\le m} \sum_{k: d(k,j)\le m}N_m(k)\\
= & \sum_k N_m(k)\sum_{j:d_{k,j}\le m}\sum_{i:d(i,j)\le m} 1\\
= & \sum_k N_m(k)\sum_{j:d_{k,j}\le m}N_m(j)\\
\le & 2\sum_k \sum_{j:d(k,j)\le m}(N_m^2(j)+N_m^2(k))\\
= & 4 \sum_j N_m^3(j)\,.
\end{align*}
The second term in \eqref{eq:tilde_gamma_rs-gamma_rs0-decomp} is bounded by, up to constant factor, $\rho^m |\mathcal U|^2\lesssim \rho^m(\sum_i N_m(i))^2$.

So that, by \Cref{ass:bound},
\begin{align}
	\mathbb E(\tilde \Gamma_{rs}-\Gamma_{rs,0})^2  &\lesssim  \sum_j N_m^3(j)+\rho^m\left[\sum_j N_m(j)\right]^2\nonumber\\
	&\lesssim n e^{2c_1m}+\rho^m n^2e^{c_1m}\,, \label{eq:tilde_gamma_rs-gamma_rs0-square-bound}
\end{align}
which implies that $\tilde \Gamma_{rs}-\Gamma_{rs,0}=o(n)$ if
$1/(2c_1)-m/\log n$ is positive and bounded away from $0$, and $c_1 <\log(\rho^{-1})$.\qed

{\bf Proof of \Cref{pro:NAR_condition}} 

Let $B=W/\tau$, then each row of $B$ has $\ell_1$ norm at most $1$.

Now
$$
\epsilon=(I-\tau B)^{-1}U=\left(\sum_{k=0}^\infty \tau^k B^k\right)U
$$
Now suppose $i$ and $j$ are two distinct nodes such that $d(i,j)=m$.

Define 
$$\tilde \epsilon_i=e_i^T(I-\tau B)^{-1}U^{(i)}$$
where
$$
U^{(i)}_j=U_j\mathds{1}_{d(i,j)\le (m-1)/2}.
$$
By construction, $\tilde\epsilon_i$ and $\tilde \epsilon_j$ are independent if $d(i,j)\ge m$.

It remains to control $\tilde\epsilon_i-\epsilon_i$. First by definition
\begin{align*}
	\tilde\epsilon_i-\epsilon_i = \sum_{k=0}^\infty \tau^k (B^k)_{i\cdot} (U^{(i)}-U)\,,
\end{align*}
where $(B^k)_{i\cdot}$ is the $i$th row of $B^k$.
Because the support of $B$ is contained in the support of $A$, thus the support of $(B^k)_{i\cdot}$ is contained in the
$k$-step neighborhood of node $i$, therefore we have $(B^k)_{i\cdot}(U^{(i)}-U)=0$ whenever $k\le (m-1)/2$.
Thus we have, for $q\ge 1$
\begin{align*}
	\|\tilde\epsilon_i-\epsilon_i\|_q =& \left\|\sum_{k>\frac{m-1}{2}} \tau^k (B^k)_{i\cdot} (U^{(i)}-U)\right\|_q\\
\le & \sup_{j}\|U_j\|_q\sum_{k>(m-1)/2}\tau^k\|B^{(k)}_{i\cdot}\|_1\\
\lesssim & \tau^{(m-1)/2}\,.
\end{align*}

Now we are ready to establish the dependence condition with $\rho=\tau^{1/2}$.

First, if $d(i,j)=m$,
\begin{align*}
	|\mathbb E(\epsilon_i\epsilon_j)|\le&|\mathbb E(\tilde\epsilon_i\tilde\epsilon_j)|+|\mathbb E(\epsilon_i-\tilde \epsilon_i,\epsilon_j)|+|\mathbb E(\tilde \epsilon_i,\epsilon_j-\tilde \epsilon_j)|\\
	\le & 0 + \|\epsilon_i-\tilde\epsilon_i\|_2\|\epsilon_j\|_2 + \|\tilde \epsilon_i\|_2\|\epsilon_j-\tilde \epsilon_j\|_2\\
	\lesssim & \tau^{(m-1)/2}
\end{align*}
and similarly,
\begin{align*}
	|\mathbb E(\epsilon_i\epsilon_j^2)|\le&|\mathbb E(\tilde \epsilon\tilde\epsilon_j^2)|+|\mathbb E(\epsilon_i-\tilde \epsilon_i)\epsilon_j^2|+|\mathbb E\tilde \epsilon_i(\epsilon_j^2-\tilde\epsilon_j^2)|\\
	\le& 0 + \|\tilde\epsilon_i-\epsilon_i\|_3\|\epsilon_j\|_3^2+ \|\tilde \epsilon_i\|_3\|(\epsilon_j\|_3+\|\tilde \epsilon_j\|_3)\|\epsilon_j-\tilde\epsilon_j\|_3\\
	\lesssim & \tau^{(m-1)/2}
\end{align*}
And if $d(\{i,j\},\{k,l\})=m$,
\begin{align*}
	|{\rm Cov}(\epsilon_i\epsilon_j,\epsilon_k\epsilon_l)|\le&|{\rm Cov}(\tilde\epsilon_i\tilde\epsilon_j,\tilde\epsilon_k\tilde\epsilon_l)|+|{\rm Cov}(\epsilon_i\epsilon_j-\tilde\epsilon_i\tilde\epsilon_j,\epsilon_k\epsilon_l)|+|{\rm Cov}(\tilde\epsilon_i\tilde\epsilon_j,\epsilon_k\epsilon_l-\tilde\epsilon_k\tilde\epsilon_l)|\\
	\lesssim& \tau^{(m-1)/2}
\end{align*}

Let $\tilde S_A=\sum_{j\in A}\tilde \xi_j$. When $d(i,A)=m$,
\begin{align*}
	|{\rm Cov}(\epsilon_i,h(t S_A))|\le &|{\rm Cov}(\tilde\epsilon_i,h(t\tilde S_A)|+|{\rm Cov}(\epsilon_i-\tilde\epsilon_i,h(t S_A))|
	+|{\rm Cov}(\tilde \epsilon_i,h(t S_A)-h(t \tilde S_A))|\\
	\le & 0 + \|\epsilon_i-\tilde \epsilon_i\|_1 \|h\|_\infty+\|\tilde \epsilon\|_2 t\|S_A-\tilde S_A\|_2 \|h'\|_\infty\\
	\lesssim & \tau^{(m-1)/2}(1+t|A|)\,.
\end{align*}
\end{document}